\documentclass{iscram}

\usepackage{graphicx}
\usepackage{float}

\usepackage{tcolorbox}
\usepackage{listings}
\usepackage{multirow}

\tcbuselibrary{most}
\usepackage{listings}

\newtcolorbox[auto counter]{myprompt}[2][]{
    title={#2}, 
    label={#1}, 
    colback=white, 
    colframe=gray!70, 
    fonttitle=\bfseries, 
    coltitle=black, 
    sharp corners, 
    boxrule=0.5mm, 
    title after break={}, 
}

\lstdefinelanguage{json}{
    basicstyle=\ttfamily\small,
    numbers=left,
    numberstyle=\scriptsize,
    stepnumber=1,
    numbersep=8pt,
    showstringspaces=false,
    breaklines=true,
    frame=lines,
    backgroundcolor=\color{gray!10},
    literate=
     *{0}{{{\color{blue}0}}}{1}
      {1}{{{\color{blue}1}}}{1}
      {2}{{{\color{blue}2}}}{1}
      {3}{{{\color{blue}3}}}{1}
      {4}{{{\color{blue}4}}}{1}
      {5}{{{\color{blue}5}}}{1}
      {6}{{{\color{blue}6}}}{1}
      {7}{{{\color{blue}7}}}{1}
      {8}{{{\color{blue}8}}}{1}
      {9}{{{\color{blue}9}}}{1}
      {:}{{{\color{red}:}}}{1}
      {,}{{{\color{red},}}}{1}
      {\{}{{{\color{black}\{}}}{1}
      {\}}{{{\color{black}\}}}}{1}
      {[}{{{\color{black}[}}}{1}
      {]}{{{\color{black}]}}}{1},
}

\usepackage{tikz}
\usetikzlibrary{shapes.geometric, arrows}

\tikzstyle{startstop} = [rectangle, rounded corners, minimum width=3cm, minimum height=1cm,text centered, draw=black, fill=red!30]
\tikzstyle{process} = [rectangle, minimum width=3cm, minimum height=1cm, text centered, draw=black, fill=blue!30]
\tikzstyle{decision} = [diamond, minimum width=3cm, minimum height=1cm, text centered, draw=black, fill=green!30]
\tikzstyle{arrow} = [thick,->,>=stealth]

\usepackage{colortbl}
\usepackage{booktabs}
\usepackage{graphicx}
\usepackage{enumitem}
\usepackage{array}
\usepackage{makecell}
\usepackage{microtype}

\usepackage{pgf}
\usepackage{pgfmath}
\usepackage{booktabs}
\usepackage{multirow}

\definecolor{high}{RGB}{255,102,0}  
\definecolor{low}{RGB}{255,235,205} 

\definecolor{high}{RGB}{200,90,0}   
\definecolor{low}{RGB}{230,230,230} 

\definecolor{high}{RGB}{100,120,180} 
\definecolor{low}{RGB}{230,230,230}  

\newcommand{\fullscorecell}[1]{%
    \pgfmathsetmacro{\rawpercent}{100*(#1 - 0.00)/1}%
    \pgfmathsetmacro{\temppercent}{max(0, min(100, \rawpercent))}%
    \xdef\percentcolor{high!\temppercent!low}%
    \cellcolor{\percentcolor}{#1}%
}

\usepackage{subcaption} 

\usepackage{array}      
\usepackage{enumitem}   
\usepackage{geometry}   

\iscramset{
  title={
    Detecting Actionable Requests and Offers on Social Media During Crises Using LLMs
  },
  short title={Detecting Actionable Requests and Offers with LLMs},
  author={
    short name={Zguir},
    full name=Ahmed El Fekih Zguir,
    affiliation={
      Qatar Computing Research Institute,\\ 
      Hamad Bin Khalifa University, Doha, Qatar\\
       \href{mailto:azguir@hbku.edu.qa}{azguir@hbku.edu.qa}
    },
  },
  author={
    full name= Ferda Ofli,
    affiliation={
      Qatar Computing Research Institute,\\ 
      Hamad Bin Khalifa University, Doha, Qatar\\
       \href{mailto:fofli@hbku.edu.qa}{fofli@hbku.edu.qa}
    },
  },
  author={
    full name= Muhammad Imran,
    affiliation={
      Qatar Computing Research Institute,\\ 
      Hamad Bin Khalifa University, Doha, Qatar\\
       \href{mailto:mimran@hbku.edu.qa}{mimran@hbku.edu.qa}
    },
  },
  CoRePaper2025={Social Media for Crisis Management}
}

\begin{document}

\maketitle

\abstract{

Natural disasters often result in a surge of social media activity, including requests for assistance, offers of help, sentiments, and general updates. To enable humanitarian organizations to respond more efficiently, we propose a fine-grained hierarchical taxonomy to systematically organize crisis-related information about requests and offers into three critical dimensions: \textit{supplies}, \textit{emergency personnel}, and \textit{actions}. Leveraging the capabilities of Large Language Models (LLMs), we introduce Query-Specific Few-shot Learning (QSF Learning) that retrieves class-specific labeled examples from an embedding database to enhance the model's performance in detecting and classifying posts. Beyond classification, we assess the actionability of messages to prioritize posts requiring immediate attention. Extensive experiments demonstrate that our approach outperforms baseline prompting strategies, effectively identifying and prioritizing actionable requests and offers.
}

\keywords{Large language models, Disaster management, Taxonomy,  Social media, Query-Specific Few-shot Learning
  }

\section{Introduction}
\label{sec:intro}


Social media platforms have become vital during crises, as they enable real-time communication and coordination during emergencies. During past events like Hurricane Dorian, Hurricane Harvey, and COVID-19, platforms like Twitter (now X) and Facebook played a crucial role, allowing people to request help and organize assistance quickly (\cite{olawale2018crisis, mihunov2020twitter}). For emergency responders and humanitarian organizations, this stream of information presents an opportunity to enhance situational awareness, allocate resources more efficiently, and provide aid where it is most needed. 
However, much of the data, including emotional expressions and commentary from those outside the affected area, is irrelevant to immediate response needs. 
To fully harness the value of social media during disasters, it is essential to develop effective methods for identifying the most relevant and actionable information amidst the noise \citep{he2017signals}.

This study focuses on identifying actionable social media posts categorized as ``requests'' (messages seeking assistance) and ``offers'' (messages providing assistance). 
Actionability, as defined by \citet{Zade2018}, refers to posts containing sufficient contextual information to assess urgency, time, and the specific location for the required assistance or available resources. Previous efforts to identify such posts have primarily focused on high-level categorization into requests and offers, with limited attention to fine-grained distinctions based on specific types of assistance. Moreover, while supervised machine learning models trained on human-labeled data are ideal for such tasks, they are often impractical in disaster scenarios due to the time, resources, and effort required for data annotation. Furthermore, adapting these models to new events or categories demands additional training data, which is challenging to obtain during emergencies.



To address these limitations, this study focuses on fine-grained identification of requests and offers without relying on extensive human-labeled data. We propose a fine-grained hierarchical taxonomy that organizes crisis-related information along three critical dimensions: \textit{supplies}, \textit{emergency personnel}, and \textit{actions}. This taxonomy was developed using a \textit{``top-down''} approach informed by documents, guidelines, and expertise from humanitarian organizations, resulting in a structured framework comprising 1,093 elements across six levels of granularity. Unlike previous \textit{``bottom-up''} approaches that rely on text mining and topic modeling techniques \citep{Durham2023}, our taxonomy ensures consistency, practicality, and alignment with real-world disaster management needs.

A significant challenge in prior research has been the oversimplification of social media classification tasks as single-label multi-class problems, which fail to account for the multifaceted nature of disaster-related posts. For instance, a tweet like, \textit{``We're cooking meals for displaced families tonight. DM me if you want to help or donate. \#CaliforniaFires,''} contains both an offer (meals for displaced families) and a request (volunteers to help). To address this, we frame the problem as a multi-label multi-class classification task, where a post can be labeled as both a request and an offer, and each post can belong to multiple other fine-grained categories.

Additionally, existing approaches often lack the granularity required for effective disaster response. For example, the category \textit{``medical supplies''} can encompass diverse items such as bandages, oxygen tanks, or mobility aids (e.g., wheelchairs),
which are critical to differentiate during a crisis. Traditional research has also predominantly focused on tangible supplies, overlooking requests for actions (e.g., search and rescue) or specific personnel (e.g., military support during riots). Our taxonomy addresses this gap by preserving granular and actionable information, enabling a more comprehensive analysis of social media posts.

To implement this taxonomy, we leverage Large Language Models (LLMs) with various prompting strategies. Additionally, we propose a Query-Specific Few-shot Learning (QSF learning) approach supported by Retrieval-Augmented Generation~\citep{lewis2020retrieval}. This method enables the classification of posts into fine-grained categories without extensive labeled data and outperforms several baselines on both real-world and synthetic data. By framing the problem as three distinct multi-label, multi-class classification tasks corresponding to supplies, actions, and emergency personnel, our framework provides a robust and scalable solution for detecting actionable requests and offers on social media data during disasters. We provide the dataset, taxonomy, and other related resources at the following URL: \url{https://crisisnlp.qcri.org/requests_offers/}.

\section{Related Work}

The growing reliance on social media during natural disasters has spurred significant research into identifying and categorizing actionable information, particularly requests and offers, to aid humanitarian efforts. Early research primarily focused on traditional machine learning techniques for classifying disaster-related posts. For instance, \citet{Purohit2014} developed and released labeled datasets and regular expressions to identify requests and offers on Twitter. Their work used cascading SVM classifiers, prioritizing precision over recall, to classify tweets into categories like money, shelter, and medical supplies. However, their approach was limited to posts containing either requests or offers, without accounting for posts that included both. Building on this, \citet{Nazer2016} incorporated additional features, such as URLs and hashtags, and used decision tree classifiers to enhance classification performance. Similarly, \citet{Devaraj2020} employed GloVe word vectors \citep{Pennington2014} to distinguish urgent posts from non-urgent ones, demonstrating the evolving sophistication of feature engineering in this domain. Taking a step further, \citet{basu2022utilizing} presented a utility-driven model for optimized resource allocation in a post-disaster scenario, based on information extracted from microblogs in real time.

The introduction of transformer architectures \citep{Vaswani2017} marked a paradigm shift in natural language processing (NLP), enabling efficient processing of long text sequences through attention mechanisms. This breakthrough paved the way for models like BERT \citep{Devlin2019} and GPT-3 \citep{Brown2020}, which established new benchmarks in text classification. Fine-tuning these pre-trained models on disaster-related tasks became a popular approach \citep{Howard2018}. For instance, \citet{Seeberger2022} fine-tuned BERT to classify disaster-related tweets into actionable categories. Prompt engineering and few-shot prompting, introduced with GPT-3, further reduced dependence on large labeled datasets, showcasing the potential for effective performance with minimal examples \citep{Brown2020}.

More recent work has leveraged multiple pre-trained transformer models to address the complexity of disaster-related tasks. For example, \citet{Zhou2022} employed BERT, RoBERTa, and XLNet to classify tweets across various disaster-related categories, outperforming traditional machine learning methods. \citet{ziaullah2024monitoring} highlighted the zero-shot capabilities of large language models (LLMs) for monitoring critical infrastructure during emergent disasters. Furthermore, \citet{lamsal2024crema} introduced crisis-specific fine-tuned transformers (\textit{CrisisTransformers}) to classify tweets into requests and offers, demonstrating significant improvements over earlier approaches.

Several studies have emphasized the importance of structured taxonomies for organizing disaster-related information. RweetMiner \citep{ullah2021rweetminer} introduced a formal framework for identifying and categorizing ``rweets'' (request tweets) into sub-types such as medical, food, and shelter, using machine learning classifiers with high precision. Similarly, \citet{basu2017resource} analyzed WhatsApp messages during the 2015 Nepal earthquake to curate resource requirements and delays, demonstrating the value of taxonomy-driven approaches for disaster preparedness. More recently, \citet{Durham2023} employed text mining and topic modeling techniques, such as latent Dirichlet allocation, to develop a \textit{bottom-up} taxonomy from tweets. In contrast, our work adopts a \textit{top-down} approach, leveraging humanitarian guidelines and expertise to define a fine-grained hierarchical taxonomy. This taxonomy captures three critical dimensions—supplies, emergency personnel, and actions—providing a robust framework for classifying posts into actionable categories.

Fine-grained classification has proven essential for improving resource allocation during crises. For instance, \citet{basu2019extracting} experimented with supervised and unsupervised models for identifying resource needs and availabilities, emphasizing the importance of granular classifications when high-quality training data is available. \citet{ullah2021rweetminer} categorized tweets into sub-types such as medical, food, and shelter using machine learning classifiers with high precision. Similarly, \citet{zhang2021topic} employed a topic model-based framework to identify the spatial distribution of demand for relief supplies, while \citet{dutt2019utilizing} proposed a methodology to match resource needs and availabilities, considering resource type, quantity, and geographical proximity. However, earlier works often simplified the problem to single-label classification, overlooking the complexity of posts containing both requests and offers. For example, a post offering food while simultaneously requesting volunteers exemplifies the need for multi-label classification.

Highlighting the need for multilingual support, \citet{vitiugin2024multilingual} introduced MulTMR, a multilingual serviceability model leveraging knowledge distillation with task-related and behavior-guided teacher models to detect and rank help requests on social media. Their approach, validated across multiple languages and disaster events, demonstrated substantial performance improvements in multilingual scenarios. Similarly, \citet{lamsal2024crema} proposed CReMa, a systematic framework integrating textual, temporal, and spatial features for cross-lingual identification and matching of requests and offers. Their multilingual embedding space and crisis-specific pretrained model significantly advanced performance benchmarks and highlighted the importance of cross-lingual analysis in disaster response.

In addition to other challenges, earlier research has also emphasized the need for actionable intelligence tailored to responders’ roles. \citet{Zade2018} highlighted issues like information overload and misinformation in integrating social media data into disaster response. They proposed shifting from general situational awareness to actionable intelligence, aligning with our redefinition of \textit{actionability}. Our approach prioritizes posts with sufficient context to drive direct actions, addressing gaps in traditional urgency-based classifications and supporting humanitarian organizations in effective decision-making during crises.

\section{Methodology}
\label{sec:method}

Our goal is to develop a robust approach capable of identifying any predefined categories of requests or offers while enabling rapid deployment without requiring large amounts of labeled data or supervised model training. To this end, we design a comprehensive taxonomy comprising three key dimensions (supplies, actions, emergency personnel). We then propose a Query-Specific Few-shot Learning (QSF learning) approach leveraging Retrieval-Augmented Generation (RAG) \citep{lewis2020retrieval} to construct few-shot prompts for message classification. Our goal is not only to improve classification performance but also to assess how these improvements generalize across different LLMs. To ensure broad applicability, we evaluate our approach using multiple instruction-tuned LLMs of varying sizes and architectures, including Llama 3 8B, Llama 3.1 8B \citep{llama}, Gemma 2 9B \citep{gemma}, Mistral 7B v0.2 \citep{mistral}, and GPT-4o mini. While all models are tested on the full range of baseline prompts and our proposed solution, GPT-4o mini—being a paid API—was evaluated exclusively on our solution (QSF Learning) to benchmark its performance relative to the other models. This diverse set of models allows us to systematically examine whether our findings generalize across LLMs with different capacities and training backgrounds. Next, we provide details of our methodology.

\subsection{Taxonomy Generation}


Most prior works, including the work by \citet{lamsal2024crema}, build on the taxonomy by \citet{Purohit2014}, which categorizes resources as tangible supplies or services requested or offered during disasters, such as monetary donations, volunteer work, shelter, clothing, and medical supplies. While valuable, these taxonomies face two key limitations: \textit{(i)} They group distinct resource types—tangible supplies (e.g., medical supplies, clothing) and intangible services (e.g., volunteer work)—failing to capture their unique characteristics and complicating accurate categorization. \textit{(ii)} They often omit critical resources frequently highlighted during disasters. For example, social media posts during Hurricane Harvey in 2017 frequently requested bottled water and baby formula, while the 2020 Beirut explosion saw significant demand for dust masks and personal protective equipment—resources absent in existing taxonomies. These gaps hinder comprehensive disaster response and underscore the need for more fine-grained categorization.



Furthermore, previous research overlooks two critical types of requests and offers: ``actions" and ``emergency personnel." Many social media posts during disasters highlight urgent actions, such as search and rescue operations, debris clearance, medical aid, food distribution, and crowd control. For example, posts during the 2015 Nepal earthquake frequently requested search and rescue teams for locating survivors \citep{bleiker2015}, while the 2023 Türkiye-Syria earthquake emphasized coordinated debris removal and emergency medical care \citep{ocha_turkey_2023}. Beyond actions, messages contain requests for trained personnel, e.g., firefighters, medical professionals, and law enforcement. For instance, the California wildfires saw appeals for firefighting reinforcements, and the Ebola outbreak required infectious disease specialists and emergency nurses.

To address these gaps, we propose a taxonomy with three main branches: \textbf{supplies}, \textbf{actions}, and \textbf{emergency personnel}. 
\begin{itemize}[noitemsep, topsep=0pt]
    \item \textbf{Supplies:} tangible resources such as medical supplies, shelter, food, water, hygiene products, and more.
    \item \textbf{Actions:} represent tasks like search and rescue, medical care, debris clearance, and food distribution.
    \item \textbf{Emergency personnel:} refers to trained responders, including paramedics, firefighters, structural engineers, and specialized volunteers.
\end{itemize}

Next, we utilized official online resources, including situation reports, guidelines, and articles from organizations such as UN OCHA, UNDP, FEMA, Red Cross, and UNHCR~\citep{UNICEF2023, UNHCR2023, OCHA2023}. 
The selection of these three dimensions—\textbf{supplies}, \textbf{actions}, and \textbf{emergency personnel}—is based on a thorough manual evaluation of the collected documents. Through careful analysis, we observed that the majority of disaster response activities naturally cluster around these three core areas: tangible resources (which we categorize as supplies), human responders and teams (personnel), and required operational activities (actions). This observation reflects the real-world practices and language used by humanitarian agencies, ensuring that our taxonomy is aligned with operational workflows and sufficiently comprehensive to capture the essential elements of disaster response.

\begin{figure}[htbp]
    \centering
    \includegraphics[width=.95\linewidth]{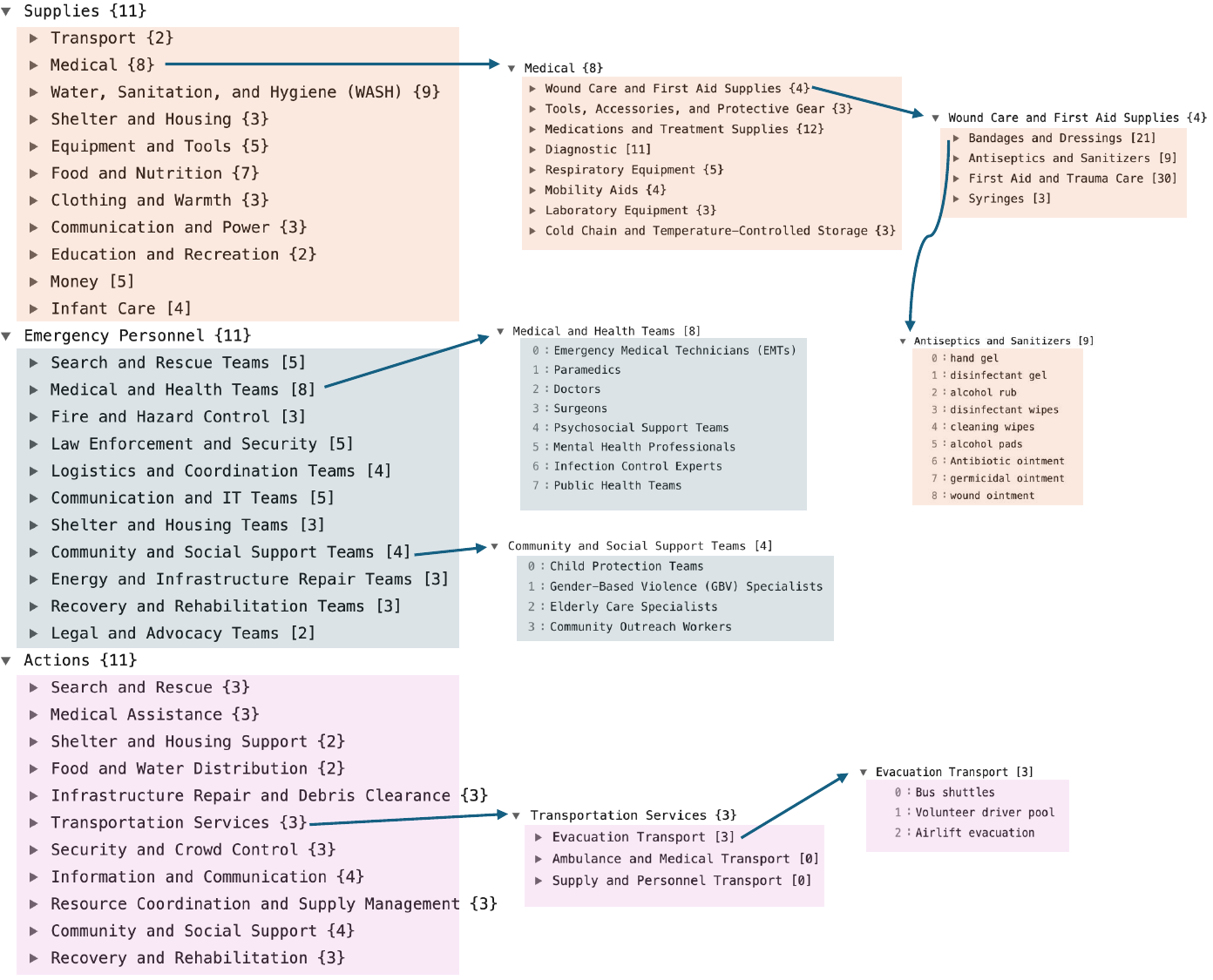} 
    \caption{Request and Offer Taxonomy: A partial representation}
    \label{fig:taxonomy}
\end{figure}

Information from 20 such sources was processed using GPT-4o, a state-of-the-art language model, to generate and augment categories within the three main branches. Table~\ref{tbl:taxonomy_prompt} outlines the prompts used to create the different taxonomy levels. The model suggested categories and sub-categories at different depths of the taxonomy underwent thorough human review, with adjustments made as needed. In the prompts, we ensure that each level provides increasingly fine-grained information about its parent category, incorporating synonyms, regional variations, and linguistic nuances. For example, the category ``bandages" was expanded to include terms such as ``Band-Aids," ``adhesive strips," and ``plasters," reflecting diverse social media expressions. The final taxonomy comprises 1,093 elements distributed as follows at different depth levels: Level-1: 3, Level-2: 33, Level-3: 129, Level-4: 635, Level-5: 271, and Level-6: 22. Figure~\ref{fig:taxonomy} shows a partial view of our taxonomy, highlighting the root branches, all categories at depth two that are directly under the root, and an expansion of some selected branches.


\begin{table}[htbp]
\scriptsize
\caption{Prompts used to generate and expand the taxonomy}
\label{tbl:taxonomy_prompt}
\begin{tabular}{|p{0.45\linewidth}|p{0.45\linewidth}|}
\hline
\textbf{Extraction Prompt: Extracting Relevant Terms} & \textbf{Level 2 Prompt: Creating the Hierarchy (Per Category)} \\
\hline
\textbf{\textless context\textgreater} \newline
Shared Context (given below) \newline
\textbf{\textless /context\textgreater} \vspace{0.5em} & 
\textbf{\textless context\textgreater} \newline
Shared Context \newline
\textbf{\textless /context\textgreater} \vspace{0.5em} \\
\textbf{\textless instruction\textgreater} \newline
You are provided with multiple documents from humanitarian organizations 
that contain terms related to disaster relief. These terms can broadly 
fall under three categories: actions, emergency personnel, and supplies.
Extract all relevant terms and group them based on which category they fall under. \newline
\textbf{\textless /instruction\textgreater} & 
\textbf{\textless instruction\textgreater} \newline
You are provided with a list of terms that fall under the category (actions, supplies, or personnel). 
Your task is to group the terms into distinct, non-overlapping groupings within this category. 
Ensure that the groupings cover all terms and are logical, clear, and comprehensive. \newline
\textbf{\textless /instruction\textgreater} \\
\hline
\textbf{Level 3 \& 4 Prompt: Refining the Combined Taxonomy} & \textbf{Levels 5 \& 6 Prompt: Expansion} \\
\hline
\textbf{\textless context\textgreater} \newline
Shared Context \newline
\textbf{\textless /context\textgreater} \vspace{0.5em} & 
\textbf{\textless context\textgreater} \newline
Shared Context \newline
\textbf{\textless /context\textgreater} \vspace{0.5em} \\
\textbf{\textless instruction\textgreater} \newline
You are provided with a taxonomy that organizes terms under actions, emergency personnel, and supplies. Your task is to refine and improve this hierarchy by reorganizing or expanding branches as needed. Different branches may vary in depth, but ensure the taxonomy remains logical, comprehensive, and well-organized. \newline
\textbf{\textless /instruction\textgreater} & 
\textbf{\textless instruction\textgreater} \newline
You are provided with a detailed taxonomy. Your task is to review and expand leaf terms by:
\begin{enumerate}[leftmargin=*]
    \item Adding synonyms for terms that are commonly referred to by different names. Make sure to include terms used both in social media and by humanitarian organizations.
    \item Adding subcategories or breaking down elements where justified.
\end{enumerate}
Only make expansions where they are necessary to improve clarity, usability, or completeness. 
Keep the taxonomy concise and avoid unnecessary additions. \newline
\textbf{\textless /instruction\textgreater} \\
\hline
\multicolumn{2}{|p{0.9\linewidth}|}{\textbf{Shared context:} In this task, you are assisting in the development of a taxonomy to classify and organize requests and offers made during disaster scenarios. The goal is to create a structured, top-down taxonomy based on a corpora of documents sourced from humanitarian organizations. These documents include situation reports, needs assessments, articles, websites, and guidelines which describe various forms of needs and offers.} \\
\hline
\end{tabular}
\end{table}

\normalsize





\subsection{Task Definition and Classifiers}

\subsubsection{Task Definition}

The goal of our task is to process an input message (e.g., a tweet) and extract key structured information relevant to crisis response. Specifically, for each input message, we aim to generate an 8-element tuple capturing its essential attributes. The tuple consists of:

\begin{itemize}[noitemsep, topsep=0pt]
    \item \textbf{Type}: A list indicating the type of message, such as \texttt{"request"}, \texttt{"offer"}, or \texttt{"other"}.
    \item \textbf{Actions (r)}: A list of requested actions (e.g., \texttt{"Search and Rescue"}).
    \item \textbf{Supplies (r)}: A list of requested supplies (e.g., \texttt{"Medical"}, \texttt{"Clothing and Warmth"}).
    \item \textbf{Personnel (r)}: A list of requested personnel (e.g., \texttt{"Medical and Health Teams"}).
    \item \textbf{Actions (o)}: A list of offered actions (e.g., \texttt{"Infrastructure Repair and Debris Clearance"}).
    \item \textbf{Supplies (o)}: A list of offered supplies (e.g., \texttt{"Money"}).
    \item \textbf{Personnel (o)}: A list of offered personnel (e.g., \texttt{"Medical and Health Teams"}).
    \item \textbf{Actionability}: A boolean value indicating whether the message contains actionable information.
\end{itemize}

Here, \texttt{(r)} denotes \textit{request} and \texttt{(o)} denotes \textit{offer}. All elements of the tuple are treated as multi-label classification tasks, except for Actionability, which is framed as a binary classification problem.

Building upon prior work~\citep{Zade2018}, we define actionability as: \textit{Information related to a crisis that either requires a response or constitutes an offer, where details such as time, location, urgency, and specific needs (e.g., quantities or actions) determine its usefulness.} For any category without relevant information, the output is either an empty list (for list-based fields) or \texttt{False} (for actionability).

Formally, let \( T \) be the set of input messages, and \( t \in T \) an individual message. We define a mapping function \( f \) as:
\[
f(t) = (\text{Type}, A_r, S_r, P_r, A_o, S_o, P_o, \text{Actionability})
\]
where:
\begin{itemize}[noitemsep, topsep=0pt]
    \item Type: A subset of \{\texttt{"request"}, \texttt{"offer"}, \texttt{"other"}\}.
    \item \( A_r \): Set of requested actions (\(A_r \subseteq A\)).
    \item \( S_r \): Set of requested supplies (\(S_r \subseteq S\)).
    \item \( P_r \): Set of requested personnel (\(P_r \subseteq P\)).
    \item \( A_o \): Set of offered actions (\(A_o \subseteq A\)).
    \item \( S_o \): Set of offered supplies (\(S_o \subseteq S\)).
    \item \( P_o \): Set of offered personnel (\(P_o \subseteq P\)).
    \item Actionability: A boolean value (\texttt{True} or \texttt{False}).
\end{itemize}

Here, \(A\), \(S\), and \(P\) represent predefined sets of target labels for actions, supplies, and personnel, respectively. Each set contains 11 distinct labels, derived from our taxonomy at depth 2.



\subsubsection{Baseline classifiers}

\begin{figure}[htbp]
    \centering
    \includegraphics[width=.77\linewidth]{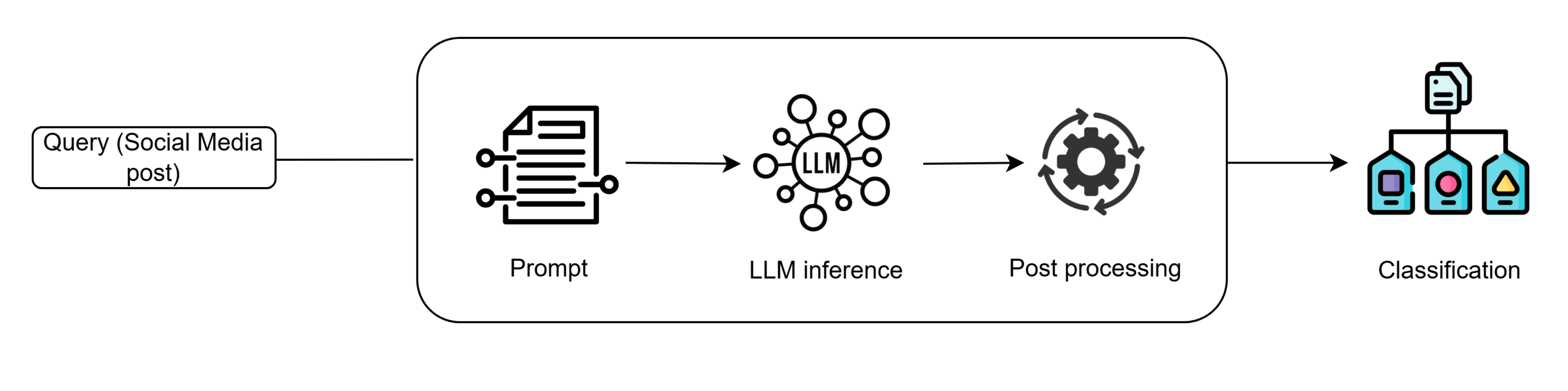} 
    \caption{Diagram of baseline (BL) classifiers. Each BL has a different prompt in the diagram.}
    \label{fig:bl_classifier}
\end{figure}

We approach the classification of crisis-related messages using LLMs, as illustrated in Figure~\ref{fig:bl_classifier}. Our method relies on prompt engineering to guide the model's output, starting with a baseline prompt and progressively refining it by incorporating additional contextual information. Each refinement results in a new classifier, designed to evaluate the impact of various prompting strategies. These classifiers differ in the level of detail they provide, specifically in terms of taxonomy depth and the inclusion of labeled examples (few-shots). The baseline prompt follows this structure:
\begin{itemize}[noitemsep, topsep=0pt]
    \item \textbf{Instruction:} In this section, we explain the LLM’s role as an advanced AI trained to classify social media posts related to crises, emphasizing the use of taxonomy and its knowledge of natural disasters.
    \item \textbf{Taxonomy:} Provides the taxonomy, defining the labels the model should output.
    \item \textbf{Context:} Discusses social media’s critical role during disasters, categorizing posts into three types: Request, Offer, and Other. It highlights the importance of identifying urgent, actionable messages, using in-context learning principles \citep{Zhou2024}.
    \item \textbf{Output Format:} Specifies the output as a JSON object with structured labels.
\end{itemize}

The baseline prompt, used with BL 1 (baseline classifier 1), can be seen below, which served as the foundation for subsequent variants detailed as follows:
\begin{itemize}[noitemsep, topsep=0pt]
    \item \textbf{BL 1 (Baseline Classifier 1):} Uses the baseline prompt with taxonomy limited to depth 2 and no few-shot examples.
    \item \textbf{BL 2:} Extends the taxonomy to depth 3, providing a more detailed hierarchical structure to improve label comprehension.
    \item \textbf{BL 3:} Incorporates few-shot prompting by adding a small set of labeled examples to the prompt, improving task understanding as shown in prior work~\citep{Brown2020}.
    \item \textbf{BL 4:} Combines BL 2 and BL 3, integrating both a detailed taxonomy and few-shot examples.
    \item \textbf{BL 5:} Builds on BL 4 by adding chain-of-thought (CoT) prompting~\citep{Kojima2022}. This classifier addresses ambiguous cases with detailed explanations and requires step-by-step reasoning for each classification decision, including actionability.
\end{itemize}

\begin{myprompt}[prompt:baseline_classification_prompt]{Baseline Prompt for Classification}
\scriptsize
\textbf{\textless Instruction\textgreater} \\
You are an advanced AI trained to label social media posts related to crisis situations, specifically natural disasters. Your goal is to label the given posts. You make use of the vast knowledge that you have about natural disasters, your knowledge of social media posts of people during those disasters, the provided taxonomy and information mentioned below to label the data.

\textbf{\textless /Instruction\textgreater}

\textbf{\textless Context\textgreater} \\
Natural disasters, such as hurricanes, wildfires, earthquakes, floods, tornadoes, landslides, etc., have significant impacts on communities. During these events, social media platforms (such as Twitter, Facebook, and Instagram) become primary channels of sharing and receiving information. This information can be broadly categorized into \textbf{Request}, \textbf{Offer}, or \textbf{Other}. Note that some posts are both \textbf{Request} and \textbf{Offer} at the same time ... (context was omitted to fit on a page)

\textbf{\textless /Context\textgreater}

\textbf{\textless Taxonomy\textgreater} \\
\{Taxonomy till depth 2 here\} \\
\textbf{\textless /Taxonomy\textgreater}

\textbf{\textless Output formats\textgreater} \\
This is the output format when the type is either \textbf{Request} or \textbf{Offer} or both:
\begin{lstlisting}[language=json, basicstyle=\ttfamily\small, frame=single]
{
    "text": "The social media post text here",
    "type": ["Request" | "Offer" | "Offer", "Request"],
    "action_request": [...],
    "personnel_request": [...],
    "supplies_request": [...],
    "action_offer": [...],
    "personnel_offer": [...],
    "supplies_offer": [...],
    "actionability": true | false
}
\end{lstlisting}

This is the output format when the type is \textbf{Other}:
\begin{lstlisting}[language=json, basicstyle=\ttfamily\small, frame=single]
{
    "text": "The social media post text here",
    "type": ["Other"]
}
\end{lstlisting}
\textbf{\textless /Output formats\textgreater}

\textbf{\textless Task\textgreater} \\
Your task is to label the following social media post based on the taxonomy and rules mentioned above. Only output the JSON dictionary and nothing else. \\
\textbf{\textless /Task\textgreater}
\end{myprompt}


\subsection{Query-Specific Few-Shot Learning Approach}
Prior studies show that including labeled examples (few-shots) in prompts improves LLM performance on classification tasks. However, in a multi-class setting, adding an equal number of shots to all classes often leads to a decline in performance, as demonstrated by \citet{imran2024evaluating}. Building on this insight, we introduce a query-specific few-shot learning strategy (QSF learning) using Retrieval-Augmented Generation (RAG)~\citep{lewis2020retrieval} to retrieve relevant, query-specific examples dynamically for each input message. 

For each message, we compute its embedding using OpenAI's \texttt{text-embedding-3-small} model and retrieve the $k/2$ most similar labeled examples from a pre-built embedding database. This database contains examples from previous crisis events. To maintain variability and prevent bias toward specific classes, we also include $k/2$ randomly selected examples. These examples are then appended to the prompt. The detailed steps of the QSF Learning algorithm can be seen in Figure~\ref{fig:rag_classifier} and are as follows:

\begin{figure}[htbp]
    \centering
    \includegraphics[width=.77\linewidth]{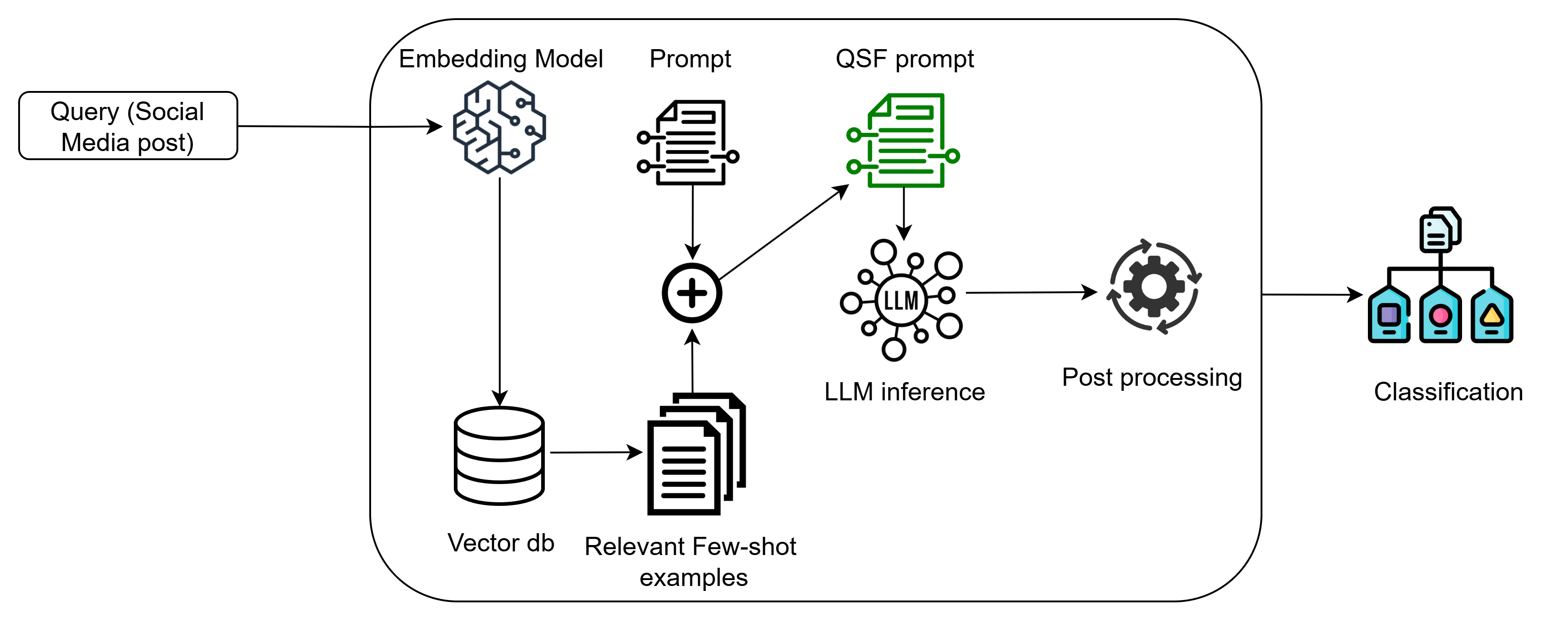} 
    \caption{A high-level overview of the Query-Specific Few-shot Learning approach}
    \label{fig:rag_classifier}
\end{figure}

\begin{enumerate}[noitemsep, topsep=0pt]
    \item \textbf{Top-k retrieval:} Embeddings for each input are computed using OpenAI's \texttt{text-embedding-3-small} model, and the $k/2$ most relevant examples are retrieved using cosine similarity.
    \item \textbf{Random sampling:} $k/2$ random examples are added for variability and to avoid over-fitting.
    \item \textbf{Mapping and appending:} Retrieved embeddings are mapped to their labeled examples and appended to the prompt.
    \item \textbf{Prompt creation:} A tailored prompt combines structured sections with retrieved examples.
    \item \textbf{Inference, cleanup, and evaluation:} The same steps as BL 1–5 are applied to process and evaluate outputs.
\end{enumerate}

\section{Data and Evaluation}
\label{sec:data_eval}

\subsection{Synthetic Data Generation}
To evaluate the proposed methodology, we sought ground-truth data aligned with our detailed taxonomy. However, to the best of our knowledge, no such fine-grained dataset currently exists. As a result, we opted to generate synthetic data (tweets) to mimic social media posts during disasters.
To achieve this, we tested various data generation methods, ultimately combining them to produce a diverse, high-quality dataset with unique elements and accurate labels. We used GPT-4o (with a May 2023 knowledge cutoff) as our data generation model.



Our initial data generation approach involved a simple prompt containing the classes from depth 2 of our taxonomy. In this approach, we asked the model to generate 100 tweets and their labels simultaneously. However, the generated data lacked naturalness and appeared robotic, as shown below:

    \textit{``Urgently need water, sanitation, and hygiene (WASH) at Westside Park. Please help!"}\\
    \textit{``Offering water, sanitation, and hygiene (WASH) at St. Andrews Church. Available anytime."}\\
    \textit{``Urgently need medical supplies at Westside Park."}

We observed that the model overly relied on the provided taxonomy labels, resulting in repetitive and unnatural outputs. Additionally, asking the model to generate 100 labeled examples in a single prompt led to repetitive text, as the model got ``lazy" and started stitching examples by merely substituting labels from the taxonomy in the same text. We observed significant improvements when we seeded the prompt with real disaster event names. This way, we ask the model to use its knowledge of the events mentioned. For instance, when focusing on the Türkiye-Syria earthquake of 2023, the generated data appeared more realistic:

    \textit{``This is so heartbreaking. Entire neighborhoods flattened in Kahramanmaraş. Stay strong. \#TurkeySyriaEarthquake"}

\begin{table}[htbp]
    \centering
    \caption{Data generation prompt evaluation results}
    \begin{tabular}{cccc}
        \toprule
        \textbf{Prompt} & \textbf{Avg. Examples} & \textbf{Mean Similarity} & \textbf{Max Similarity} \\
        \midrule
        P1 & 38 & 0.4169 & 0.8154 \\
        P2 & 40 & 0.4022 & 0.7617 \\
        P3 & 68 & 0.3660 & 0.8189 \\
        P4 & 70 & 0.3727 & 0.8302 \\
        \bottomrule
    \end{tabular}
    \label{tab:prompt_evaluation_results}
\end{table}

We further redesigned the prompt structure to address other issues, e.g., related to the length of the tweets, realisticness, etc., through iterative rounds of prompt engineering. The final structure included detailed instructions, context, output format, and enhanced instructions, formatted as follows:

\begin{myprompt}[prompt:synthetic_data_generation]{Synthetic Data Generation Prompt}
\scriptsize
\textbf{\textless Instruction\textgreater} \\
You are an advanced AI trained to generate realistic and diverse synthetic social media posts related to crisis situations, specifically natural disasters. Your goal is to create posts that closely mimic real-life data while ensuring creativity, uniqueness, and variability. You make use of the vast knowledge that you have about natural disasters and social media posts of people during those disasters to generate the data. \\
\textbf{\textless /Instruction\textgreater}

\textbf{\textless Context\textgreater} \\
\{same context as baseline prompt for classification\} \\
\textbf{\textless /Context\textgreater}

\textbf{\textless EnhancedInstructions\textgreater} \\
The data that you generate should adhere to the following guidelines:
\begin{itemize}
    \item All generated data must be unique. Avoid creating similar posts with minor variations (e.g., changing only the place or disaster name).
    \item Data should look as realistic as possible, simulating posts created by humans during a natural disaster.
    \item Posts should be diverse, including noisy data, partial data, and complete data.
    \item Incorporate realistic elements such as links, phone numbers, hashtags, grammatical mistakes, abbreviations, etc., to enhance authenticity.
    \item Integrate references to actual past events to ground the data in reality. Use the following disasters as reference points: \{list of chosen disasters\}.
    \item Ensure a balance of content types (\textbf{Request}, \textbf{Offer}, and \textbf{Other}) in the generated posts.
\end{itemize}
\textbf{\textless /EnhancedInstructions\textgreater}

\textbf{\textless OutputFormat\textgreater} \\
The output should be a JSON object containing 100 generated posts in the following format:
\begin{lstlisting}[language=json, basicstyle=\ttfamily\small, frame=single]
{
    "posts": [
        "generated text 1 here",
        "generated text 2 here",
        ...
        "generated text n"
    ]
}
\end{lstlisting}
\textbf{\textless /OutputFormat\textgreater}

\textbf{\textless Task\textgreater} \\
Your task is to generate $N$ realistic social media posts based on the context and instructions provided above. Use the same context as the baseline prompt for classification. Each post should adhere to the guidelines, reflect a variety of disaster-related content, and maintain uniqueness. \\
\textbf{\textless /Task\textgreater}
\end{myprompt}

To test the best data generation approach through prompting, in total, we created four prompts based on the above-mentioned structure, as follows:
\begin{itemize}[noitemsep, topsep=0pt]
    \item \textbf{Prompt 1:} Taxonomy + labeling (data and labels generated together).
    \item \textbf{Prompt 2:} No taxonomy + labeling.
    \item \textbf{Prompt 3:} Taxonomy (data generation only).
    \item \textbf{Prompt 4:} No taxonomy (data generation only).
\end{itemize}


We instructed the LLM to generate 75 examples per prompt, using a temperature setting of 0.8 to enhance variability. Each prompt was executed 10 times to evaluate which prompting strategy produced better results, particularly in terms of generating more unique tweets. We calculated the mean and maximum similarity of the generated texts and recorded the number of messages produced. Specifically, we calculated the cosine similarity between all pairs of embeddings of generated examples. OpenAI’s \texttt{text-embedding-3-small} model was used to create the embeddings. While the model was instructed to generate 75 examples, the output varied across runs. Table~\ref{tab:prompt_evaluation_results} summarizes the averaged metrics across the 10 iterations.

%
%
%
%

The results indicate that generating and labeling simultaneously (Prompts 1 and 2) produces fewer examples due to token limitations and results in higher similarity between outputs. This aligns with prior research showing that breaking complex tasks into smaller chunks improves the performance of LLMs \citep{khot2023}.

Based on these findings, we selected Prompt 4 (no taxonomy, data generation only) for 80\% of the data and Prompt 3 (taxonomy, data generation only) for 20\% to balance realism and coverage. We focused on disasters such as the Pakistan floods (2014), Türkiye-Syria earthquakes (2023), Australian bushfires (2019), Hurricane Maria (2017), and Haiti earthquake (2010). From an initial pool of 2,000 generated examples, duplicates (pairs with cosine similarity $> 0.925$) and unnatural outputs were removed, resulting in 1,346 high-quality examples.
\begin{figure}[H]
    \centering
    \begin{subfigure}[b]{0.38\textwidth}
        \centering
        \includegraphics[width=\textwidth]{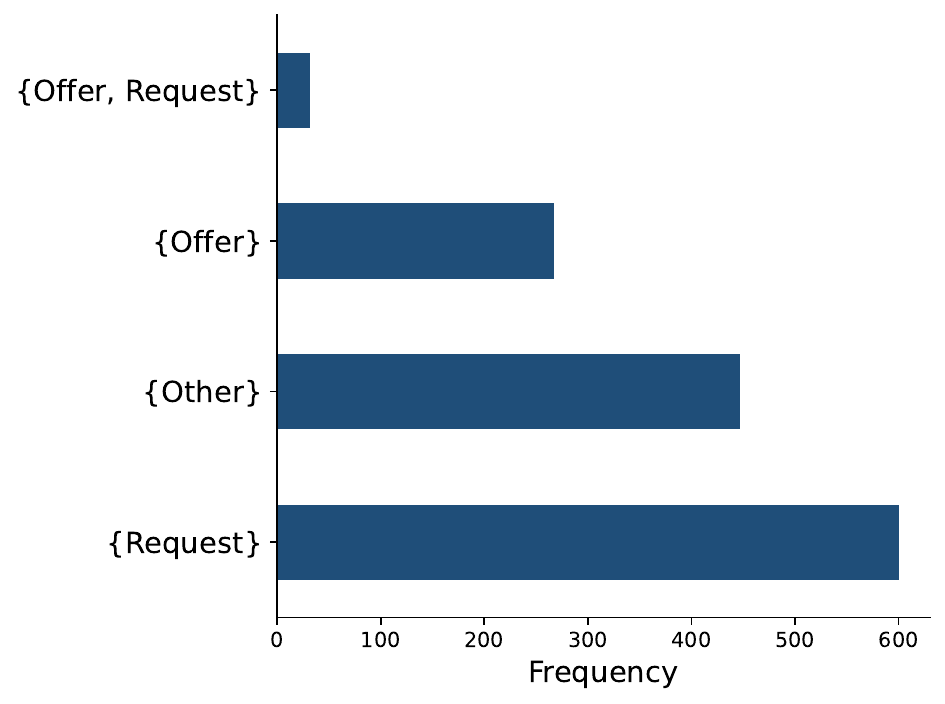}
        \caption{\textbf{Type}}
        \label{fig:type_dis}
    \end{subfigure}
    \begin{subfigure}[b]{0.40\textwidth}
        \centering
        \includegraphics[width=\textwidth]{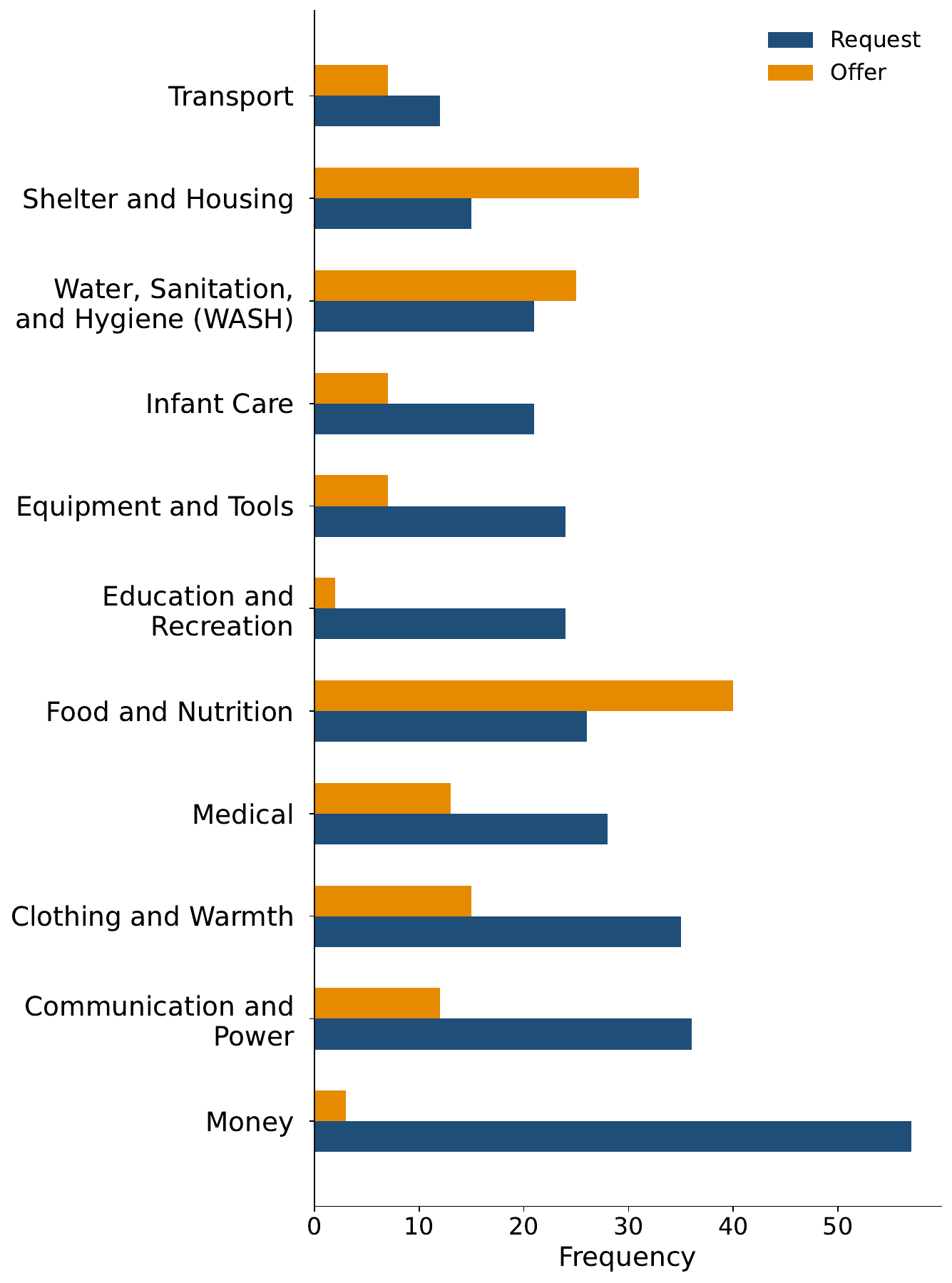}
        \caption{Supplies}
        \label{fig:supplies_dis}
    \end{subfigure}
    \caption{Distribution of the synthetic generated data across (a) Type and (b) Supply categories}
    \label{fig:data_types_supplies}
\end{figure}

\begin{figure}[htbp]
    \centering
    \begin{subfigure}[b]{0.45\textwidth}
        \centering
        \includegraphics[width=\textwidth, trim={0 0 0 0}, clip]{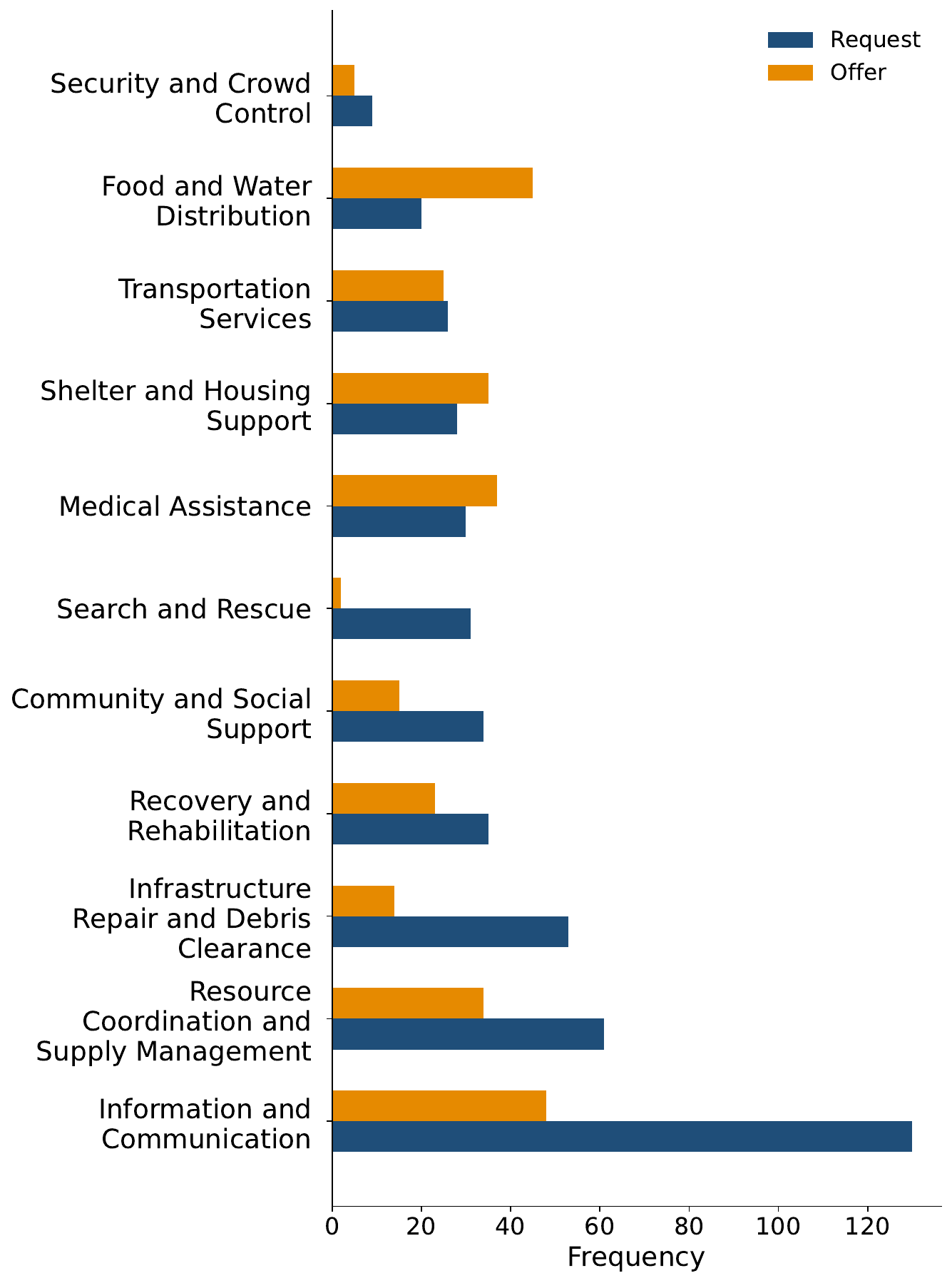}
        \caption{\textbf{Actions}}
        \label{fig:action_dis}
    \end{subfigure}
    \hfill
    \begin{subfigure}[b]{0.45\textwidth}
        \centering
        \includegraphics[width=\textwidth, trim={0 0 0 0}, clip]{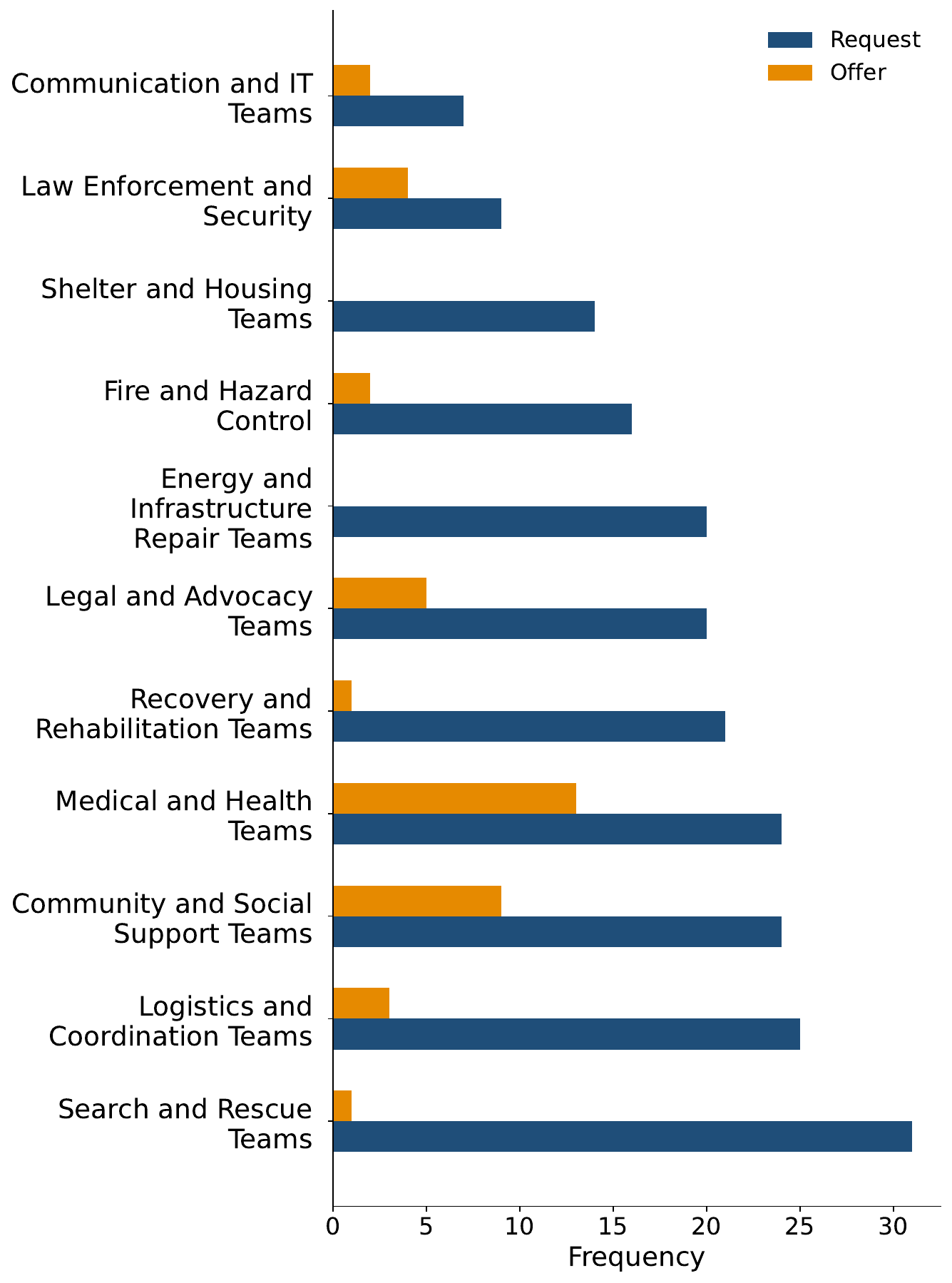}
        \caption{\textbf{Emergency Personnel}}
        \label{fig:personnel_dis}
    \end{subfigure}
    \caption{Distribution of the synthetic generated data across (a) Action and (b) Emergency Personnel categories}
    \label{fig:data_actions_personnel}
\end{figure}

The generated data was labeled using few-shot learning, guided by human-labeled examples. Human annotators reviewed 40\% of the data (538 examples) and found fewer than 10\% mislabeled, which were manually corrected. Specifically, for the message ``type" task, only 10 examples were mislabeled, and for other categories (\textit{actions, supplies, personnel}), 50 examples had minor issues with extra or missing labels.
%
%
Posts labeled as type ``Other" (neither a request nor an offer) were excluded from further labeling for actions, supplies, personnel, and actionability categories. Figures~\ref{fig:data_types_supplies} \& \ref{fig:data_actions_personnel} show the distribution of the generated data for the message types, supplies, actions, and emergency personnel for both requests and offers. As for actionability, from the 899 posts that were either requests or offers, 748 were actionable while 151 were not actionable.

\subsection{Real-World Data}
While synthetic data enabled us to create a controlled and balanced dataset aligned with our detailed taxonomy, it was essential to evaluate the robustness of our approach on real-world data, where posts are inherently noisier, less structured, and often incomplete. To this end, we utilized an existing dataset of disaster-related tweets collected and originally annotated by \citet{Purohit2014} and later improved by \citet{lamsal2024crema}. Their dataset contains tweets labeled as either requests or offers during the Hurricane Sandy disaster, providing a solid foundation for real-world evaluation.

From this dataset, we randomly sampled 300 tweets and manually annotated them according to our taxonomy. Each tweet was reviewed to assign the relevant labels across all applicable categories, including type, supplies, actions, personnel, and actionability.
Importantly, this subset of real-world data allowed us to test the model's performance in scenarios where noise and incomplete information are prevalent. For example, consider the following tweet:

\begin{quote} \textit{\#LiveWire Game Donated \$10,000 to Hurricane Sandy Voters: The rapper wanted to help storm victi... http://t.co/EWkLK4ph \#LiveWireRecords} \end{quote}

Here, part of the message is truncated, and the text lacks explicit mentions of key elements such as location or clear action verbs—common challenges encountered in authentic social media posts. By incorporating such examples, we ensured that the model's ability to generalize extends beyond synthetic, well-formed data and can handle the ambiguity and noise characteristic of real-world platforms like Twitter.

Due to the labor-intensive nature of manual labeling, especially given the multi-label setup and the breadth of categories, we limited the annotation to 300 examples. Nevertheless, this subset proved sufficient to validate that our methodology remains effective even when applied to naturally occurring, less structured data.

\subsubsection{Evaluation metrics}

We split the synthetic dataset into 50\% training (673 examples) and 50\% evaluation (673 examples). Examples used for few-shot prompting in BL 3, 4, and 5 are strictly from the training set. For the QSF Learning method, relevant examples are retrieved exclusively from the training set to ensure no data leakage into the evaluation set. For the real-world dataset, we use a training set of 107 examples and a test set of 200 examples.
We evaluate the classification performance using micro F1-scores across all multi-label tasks: type, supplies, actions, and personnel. Micro averaging aggregates contributions of all classes globally, treating each instance-label pair equally, making it particularly well-suited for multi-label classification tasks.
For the binary classification task of actionability, we report macro F1-scores to account for both classes equally, regardless of class distribution.

\section{Results and Discussion}

\begin{figure}[htbp]
    \centering
    \includegraphics[width=0.7\textwidth]{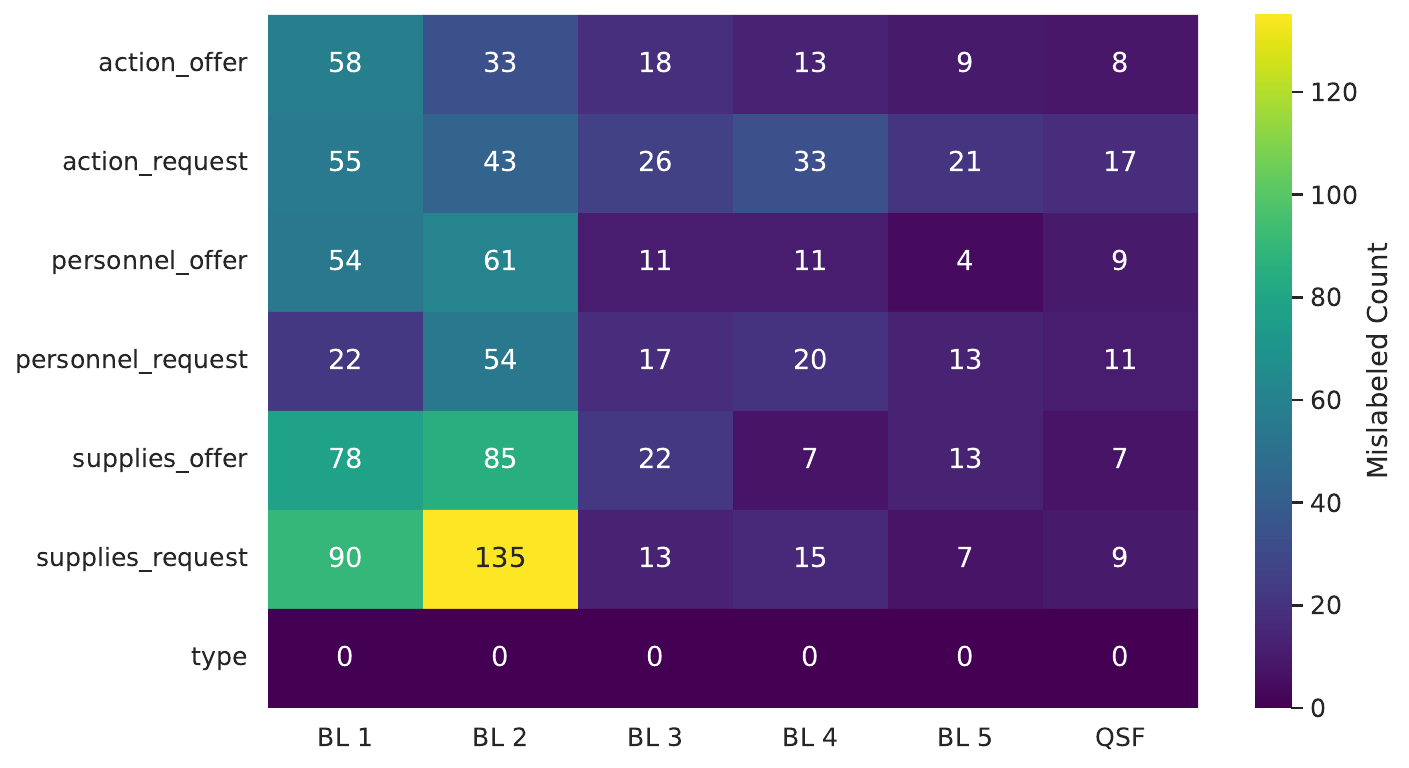} 
    \caption{Error distribution during inference on synthetic data using Llama 3}
    \label{fig:classification_errors}
\end{figure}

Our task includes one binary classification task (i.e., actionability) and seven multi-label classification tasks. Despite careful prompt design, the model occasionally produces problematic outputs, such as assigning labels outside the taxonomy.
Figure~\ref{fig:classification_errors} shows the distribution of the errors from models (Llama 3) at the classification/inference time. 
BL 1 and 2 exhibit a substantial number of errors (359 and 410, respectively), whereas BL 3 through 6 show a significant reduction. This sharp decline indicates that adding few-shot examples to the prompts significantly enhances the LLM's ability to generate taxonomy-compliant outputs. BL 3 and 4, which include few-shot examples, produce similar counts, as do BL 5 and our QSF learning technique. The difference between BL 4 and 5 can be attributed to the inclusion of chain-of-thought prompting in BL 5, which encourages step-by-step reasoning and results in better alignment with the taxonomy.

In addition to errors, the model sometimes provides broken responses that cannot be processed and evaluated even after post processing. Only BL 3 and 4 result in such responses, with BL 3 resulting in 38 broken responses, and BL 4 resulting in 8 broken responses. Both prompts lack chain-of-thought prompting, indicating that while few-shot examples reduce mislabels, incorporating structured reasoning further stabilizes the output quality.

Table~\ref{tab:synthetic_results} and ~\ref{tab:real_results} present evaluation results for our multi-task, multi-label classification tasks on both synthetic and real-world data. We run our evaluation on multiple models, namely: Llama 3 8B, Llama 3.1 8B, Gemma 2 9B, Mistral 7B v0.2, and GPT-4o mini. For the \textit{Type} task, baseline prompts (BL1–BL5) exhibit gradual improvements, with BL5 consistently outperforming earlier versions across all models. This is likely due to the relative simplicity of the \textit{Type} task and its limited label space (\textit{request}, \textit{offer}, \textit{other}). However, incorporating QSF learning yields the highest F1-scores across most models. For instance, on synthetic data, Llama 3.1 improves from 0.86 (BL5) to 0.89, while GPT-4o mini achieves a peak F1-score of 0.92. On real-world data, although absolute F1-scores are lower due to increased complexity, the same pattern persists. Models such as Gemma 2 and Llama 3 benefit from a 2-3\% improvement over their best baselines, further confirming the advantage of QSF Learning for this task.

\begin{table}[ht]
\centering
\begin{tabular}{llcccccc}
\toprule
\textbf{Task} & \textbf{Model} & \textbf{BL1} & \textbf{BL2} & \textbf{BL3} & \textbf{BL4} & \textbf{BL5} & \textbf{QSF} \\
\midrule

\multirow{5}{*}{\textbf{Type}} 
  & Mistral      & \fullscorecell{0.73} & \fullscorecell{0.74} & \fullscorecell{0.80} & \fullscorecell{0.81} & \fullscorecell{0.87} & \fullscorecell{0.87} \\
  & Llama 3.1    & \fullscorecell{0.71} & \fullscorecell{0.72} & \fullscorecell{0.82} & \fullscorecell{0.82} & \fullscorecell{0.86} & \fullscorecell{0.89} \\
  & Gemma 2      & \fullscorecell{0.73} & \fullscorecell{0.74} & \fullscorecell{0.85} & \fullscorecell{0.83} & \fullscorecell{0.88} & \fullscorecell{0.86} \\
  & Llama 3      & \fullscorecell{0.75} & \fullscorecell{0.77} & \fullscorecell{0.84} & \fullscorecell{0.82} & \fullscorecell{0.85} & \fullscorecell{0.89} \\
  & GPT-4o mini  &  --  &  --  &  --  &  --  &  --  & \fullscorecell{0.92} \\
\midrule

\multirow{5}{*}{\textbf{Supplies}} 
  & Mistral      & \fullscorecell{0.62} & \fullscorecell{0.57} & \fullscorecell{0.79} & \fullscorecell{0.82} & \fullscorecell{0.82} & \fullscorecell{0.81} \\
  & Llama 3.1    & \fullscorecell{0.57} & \fullscorecell{0.58} & \fullscorecell{0.75} & \fullscorecell{0.77} & \fullscorecell{0.82} & \fullscorecell{0.84} \\
  & Gemma 2      & \fullscorecell{0.69} & \fullscorecell{0.63} & \fullscorecell{0.79} & \fullscorecell{0.75} & \fullscorecell{0.75} & \fullscorecell{0.79} \\
  & Llama 3      & \fullscorecell{0.38} & \fullscorecell{0.41} & \fullscorecell{0.76} & \fullscorecell{0.78} & \fullscorecell{0.79} & \fullscorecell{0.79} \\
  & GPT-4o mini  &  --  &  --  &  --  &  --  &  --  & \fullscorecell{0.88} \\
\midrule

\multirow{5}{*}{\textbf{Actions}} 
  & Mistral      & \fullscorecell{0.49} & \fullscorecell{0.51} & \fullscorecell{0.55} & \fullscorecell{0.57} & \fullscorecell{0.62} & \fullscorecell{0.77} \\
  & Llama 3.1    & \fullscorecell{0.47} & \fullscorecell{0.48} & \fullscorecell{0.55} & \fullscorecell{0.59} & \fullscorecell{0.71} & \fullscorecell{0.77} \\
  & Gemma 2      & \fullscorecell{0.49} & \fullscorecell{0.53} & \fullscorecell{0.61} & \fullscorecell{0.61} & \fullscorecell{0.69} & \fullscorecell{0.73} \\
  & Llama 3      & \fullscorecell{0.40} & \fullscorecell{0.49} & \fullscorecell{0.54} & \fullscorecell{0.57} & \fullscorecell{0.64} & \fullscorecell{0.75} \\
  & GPT-4o mini  &  --  &  --  &  --  &  --  &  --  & \fullscorecell{0.77} \\
\midrule

\multirow{5}{*}{\textbf{Personnel}} 
  & Mistral      & \fullscorecell{0.44} & \fullscorecell{0.40} & \fullscorecell{0.46} & \fullscorecell{0.49} & \fullscorecell{0.54} & \fullscorecell{0.64} \\
  & Llama 3.1    & \fullscorecell{0.45} & \fullscorecell{0.41} & \fullscorecell{0.51} & \fullscorecell{0.52} & \fullscorecell{0.62} & \fullscorecell{0.70} \\
  & Gemma 2      & \fullscorecell{0.51} & \fullscorecell{0.53} & \fullscorecell{0.57} & \fullscorecell{0.56} & \fullscorecell{0.59} & \fullscorecell{0.66} \\
  & Llama 3      & \fullscorecell{0.31} & \fullscorecell{0.35} & \fullscorecell{0.46} & \fullscorecell{0.47} & \fullscorecell{0.54} & \fullscorecell{0.69} \\
  & GPT-4o mini  &  --  &  --  &  --  &  --  &  --  & \fullscorecell{0.72} \\
\midrule

\multirow{5}{*}{\textbf{Actionability}} 
  & Mistral      & \fullscorecell{0.64} & \fullscorecell{0.63} & \fullscorecell{0.70} & \fullscorecell{0.67} & \fullscorecell{0.72} & \fullscorecell{0.67} \\
  & Llama 3.1    & \fullscorecell{0.54} & \fullscorecell{0.56} & \fullscorecell{0.50} & \fullscorecell{0.44} & \fullscorecell{0.54} & \fullscorecell{0.59} \\
  & Gemma 2      & \fullscorecell{0.62} & \fullscorecell{0.64} & \fullscorecell{0.68} & \fullscorecell{0.67} & \fullscorecell{0.78} & \fullscorecell{0.49} \\
  & Llama 3      & \fullscorecell{0.68} & \fullscorecell{0.60} & \fullscorecell{0.60} & \fullscorecell{0.64} & \fullscorecell{0.58} & \fullscorecell{0.81} \\
  & GPT-4o mini  &  --  &  --  &  --  &  --  &  --  & \fullscorecell{0.60} \\
\bottomrule
\end{tabular}
\caption{Results (F1-scores) for all models and baselines on synthetic data.}
\label{tab:synthetic_results}
\end{table}

\begin{table}[ht]
\centering
\begin{tabular}{llcccccc}
\toprule
\textbf{Task} & \textbf{Model} & \textbf{BL1} & \textbf{BL2} & \textbf{BL3} & \textbf{BL4} & \textbf{BL5} & \textbf{QSF} \\
\midrule

\multirow{5}{*}{\textbf{Type}} 
  & Mistral      & \fullscorecell{0.51} & \fullscorecell{0.56} & \fullscorecell{0.63} & \fullscorecell{0.65} & \fullscorecell{0.69} & \fullscorecell{0.74} \\
  & Llama 3.1    & \fullscorecell{0.64} & \fullscorecell{0.66} & \fullscorecell{0.74} & \fullscorecell{0.73} & \fullscorecell{0.72} & \fullscorecell{0.75} \\
  & Gemma 2      & \fullscorecell{0.55} & \fullscorecell{0.58} & \fullscorecell{0.69} & \fullscorecell{0.67} & \fullscorecell{0.74} & \fullscorecell{0.77} \\
  & Llama 3      & \fullscorecell{0.54} & \fullscorecell{0.54} & \fullscorecell{0.75} & \fullscorecell{0.73} & \fullscorecell{0.74} & \fullscorecell{0.77} \\
  & GPT-4o mini  &  --  &  --  &  --  &  --  &  --  & \fullscorecell{0.72} \\
\midrule

\multirow{5}{*}{\textbf{Supplies}} 
  & Mistral      & \fullscorecell{0.58} & \fullscorecell{0.53} & \fullscorecell{0.83} & \fullscorecell{0.84} & \fullscorecell{0.79} & \fullscorecell{0.83} \\
  & Llama 3.1    & \fullscorecell{0.41} & \fullscorecell{0.45} & \fullscorecell{0.63} & \fullscorecell{0.69} & \fullscorecell{0.80} & \fullscorecell{0.82} \\
  & Gemma 2      & \fullscorecell{0.76} & \fullscorecell{0.57} & \fullscorecell{0.80} & \fullscorecell{0.79} & \fullscorecell{0.71} & \fullscorecell{0.85} \\
  & Llama 3      & \fullscorecell{0.41} & \fullscorecell{0.32} & \fullscorecell{0.79} & \fullscorecell{0.81} & \fullscorecell{0.81} & \fullscorecell{0.83} \\
  & GPT-4o mini  &  --  &  --  &  --  &  --  &  --  & \fullscorecell{0.85} \\
\midrule

\multirow{5}{*}{\textbf{Actions}} 
  & Mistral      & \fullscorecell{0.21} & \fullscorecell{0.22} & \fullscorecell{0.37} & \fullscorecell{0.36} & \fullscorecell{0.43} & \fullscorecell{0.54} \\
  & Llama 3.1    & \fullscorecell{0.23} & \fullscorecell{0.21} & \fullscorecell{0.36} & \fullscorecell{0.39} & \fullscorecell{0.38} & \fullscorecell{0.55} \\
  & Gemma 2      & \fullscorecell{0.34} & \fullscorecell{0.31} & \fullscorecell{0.48} & \fullscorecell{0.49} & \fullscorecell{0.50} & \fullscorecell{0.50} \\
  & Llama 3      & \fullscorecell{0.29} & \fullscorecell{0.26} & \fullscorecell{0.41} & \fullscorecell{0.43} & \fullscorecell{0.42} & \fullscorecell{0.57} \\
  & GPT-4o mini  &  --  &  --  &  --  &  --  &  --  & \fullscorecell{0.61} \\
\midrule

\multirow{5}{*}{\textbf{Personnel}} 
  & Mistral      & \fullscorecell{0.07} & \fullscorecell{0.08} & \fullscorecell{0.00} & \fullscorecell{0.19} & \fullscorecell{0.21} & \fullscorecell{0.21} \\
  & Llama 3.1    & \fullscorecell{0.06} & \fullscorecell{0.07} & \fullscorecell{0.19} & \fullscorecell{0.17} & \fullscorecell{0.27} & \fullscorecell{0.47} \\
  & Gemma 2      & \fullscorecell{0.04} & \fullscorecell{0.14} & \fullscorecell{0.22} & \fullscorecell{0.19} & \fullscorecell{0.25} & \fullscorecell{0.38} \\
  & Llama 3      & \fullscorecell{0.06} & \fullscorecell{0.00} & \fullscorecell{0.19} & \fullscorecell{0.22} & \fullscorecell{0.22} & \fullscorecell{0.31} \\
  & GPT-4o mini  &  --  &  --  &  --  &  --  &  --  & \fullscorecell{0.35} \\
\midrule

\multirow{5}{*}{\textbf{Actionability}} 
  & Mistral      & \fullscorecell{0.60} & \fullscorecell{0.53} & \fullscorecell{0.49} & \fullscorecell{0.45} & \fullscorecell{0.47} & \fullscorecell{0.58} \\
  & Llama 3.1    & \fullscorecell{0.43} & \fullscorecell{0.47} & \fullscorecell{0.47} & \fullscorecell{0.40} & \fullscorecell{0.60} & \fullscorecell{0.66} \\
  & Gemma 2      & \fullscorecell{0.50} & \fullscorecell{0.51} & \fullscorecell{0.69} & \fullscorecell{0.64} & \fullscorecell{0.63} & \fullscorecell{0.70} \\
  & Llama 3      & \fullscorecell{0.45} & \fullscorecell{0.48} & \fullscorecell{0.52} & \fullscorecell{0.54} & \fullscorecell{0.45} & \fullscorecell{0.47} \\
  & GPT-4o mini  &  --  &  --  &  --  &  --  &  --  & \fullscorecell{0.72} \\
\bottomrule
\end{tabular}
\caption{Results (F1-scores) for all models and baselines on real-world data.}
\label{tab:real_results}
\end{table}

For the \textit{Supplies} task, baseline prompts improve steadily from BL1 to BL5, particularly when few-shot prompting is introduced. Given the task's complexity, involving 11 labels and 2048 label combinations ($2^{11}$), static prompts alone achieve moderate performance. The QSF Learning consistently leads to the highest scores across both synthetic and real-world data. For example, in the synthetic setting, Llama 3.1 improves from 0.82 (BL5) to 0.84, while GPT-4o mini achieves an F1 score of 0.88. A similar trend is observed in real data, where the QSF classifier (``QSF") consistently boosts model performance to around 0.83–0.85, suggesting that context-aware prompting is particularly effective in providing the necessary contextual variety to handle high-variability multi-label tasks.

The \textit{Actions} task follows a similar trajectory. Baseline prompts struggle, especially in real data, with BL1 and BL2 achieving F1 scores as low as 0.20–0.30 across models. However, each subsequent enhancement results in gradual gains. The most notable improvements are observed with QSF learning, which increases F1 scores by up to 15\% compared to the strongest static baseline. For instance, on synthetic data, Mistral improves from 0.62 (BL5) to 0.77, while on real-world data, Llama 3's performance increases from 0.42 (BL5) to 0.57. This indicates that tailoring examples dynamically allows the model to better capture complex action patterns.

The \textit{Personnel} task presents significant challenges across both datasets, reflected by generally lower F1 scores. In real-world data, baseline prompts achieve particularly poor performance, often under 0.30. Despite this, our QSF learning approach consistently yields the highest improvements. For example, Llama 3.1 improves from 0.27 (BL5) to 0.47, and Gemma 2 increases from 0.25 to 0.38. On synthetic data, the trend is similar, with performance gains of approximately 10–15\% across models. These results suggest that dynamically selecting query-relevant few-shot examples is crucial for addressing class imbalance and label sparsity in the \textit{Personnel} task.

Examining the detailed taxonomy reveals that \textit{supplies} are the most granular dimension, followed by \textit{actions} and \textit{personnel}. This reflects the humanitarian organization corpora used to build the taxonomy, which primarily emphasize \textit{supplies}. The taxonomy itself contains 946 supply-related elements, compared to 90 for actions and 57 for personnel, contributing to richer coverage in that dimension.

Furthermore, supplies are typically mentioned explicitly in social media posts, making them easier to classify. In contrast, references to actions or personnel are often implicit, requiring the model to infer intent, which increases prediction difficulty. This challenge is compounded by class prevalence: supplies dominate both our synthetic and real datasets, while actions and personnel are less frequently labeled.

In our real-world dataset (307 labeled posts: 107 for retrieval, 200 for testing), posts labeled with supplies significantly outnumber those mentioning actions or personnel. Consequently, the embedding database used in our QSF classifier is denser and more representative for supplies, providing the model with more relevant examples. These factors—the taxonomy’s granularity, explicit mentions, and higher class frequency—collectively explain the stronger classification performance observed for supplies.

The \textit{Actionability} task shows mixed results. On synthetic data, the improvements due to our QSF learning approach vary by model. For instance, Llama 3.1 sees a moderate increase from 0.54 (BL5) to 0.59, while Llama 3 shows a significant gain from around 0.58 to 0.81. However, Gemma 2’s QSF score (0.49) is lower than its best baseline (0.78), indicating that the benefits of QSF learning might depend on how well the model’s underlying representation aligns with the task. In real-world data, though, our method generally yields better F1-scores i.e., 0.66 for Llama 3.1 and 0.70 for Gemma 2, which points to its potential to enhance recall for less frequent classes.

These findings reveal key insights into both model behavior and prompt design in multi-task, multi-label classification. Across both synthetic and real-world datasets, QSF learning consistently delivers the most substantial performance gains, particularly for tasks with complex label structures such as \textit{Actions} and \textit{Personnel}. By tailoring contextual examples to each input, this technique enables models to better handle label imbalance and variability, leading to superior generalization. While static prompt enhancements (BL1–BL5) result in incremental improvements, they plateau quickly, especially in granular tasks. Interestingly, the results show that QSF learning not only improves performance across tasks but also narrows the gap between models of different sizes and capabilities. High-performing models like GPT-4o mini benefit from this technique, but smaller models such as Llama 3.1 and Gemma 2 also achieve competitive results when QSF learning is applied. This indicates that prompt engineering, particularly dynamic few-shot approaches, plays a pivotal role in bridging the performance disparity between models, allowing even smaller models to generalize effectively in complex, real-world scenarios.
\section{Limitations and Future work}
\label{sec:limitations}

In this section, we outline the key limitations and biases of our study. One primary source of bias stems from the dataset used. While a portion of our evaluation is conducted on a real-world dataset comprising 300 manually labeled tweets, the relatively small size of this dataset limits its ability to fully capture the variability, noise, and evolving nature of social media posts during crises. To address the scarcity of large-scale curated datasets, we also utilized synthetically generated data produced using GPT-4o. Although GPT-4o provides diverse and partially grounded examples, it may not entirely replicate the informal language, emerging trends, or unpredictability present in real-world streams.

Another limitation is the focus solely on English-language posts, which excludes the multilingual nature of crisis communication in many global contexts. Expanding to multilingual datasets is an important direction for enhancing the generalizability of our approach.

Additionally, while the taxonomy introduces valuable structure and granularity to the classification process, its evaluation has primarily been qualitative. A more systematic, quantitative assessment of the taxonomy’s depth, completeness, and impact on classification performance would provide deeper insights and potentially inform refinements to better support crisis response applications.

For future work, we recommend expanding the real-world dataset, incorporating multilingual posts, and exploring alternative data generation strategies to improve robustness. Further, developing quantitative metrics to assess the taxonomy's design and conducting controlled experiments to evaluate its effect on classification outcomes will be key to enhancing its utility. Incorporating optimization techniques, such as fine-tuning or multi-agent frameworks, may also offer additional performance gains.
\section{Conclusion}
In this work, we introduced a fine-grained hierarchical taxonomy and a dynamic few-shot prompting technique to improve the detection of actionable requests and offers in social media posts during natural disasters. Our taxonomy organizes crisis-related information related to requests and offers into three core dimensions: \textit{supplies}, \textit{emergency personnel}, and \textit{actions}. By leveraging the capabilities of Large Language Models, we demonstrated through extensive experiments that our approach significantly outperforms baseline prompting methods in accurately identifying and prioritizing actionable content. These contributions provide a valuable framework for enhancing the efficiency of humanitarian organizations in crisis management and rapid response. Future work will focus on expanding the approach to diverse disaster scenarios, integrating real-world data, and incorporating advancements in next-generation LLMs to further refine performance and adaptability.

\bibliographystyle{plainnat} 
\bibliography{references}   

\end{document}